
\tolerance=1200
\magnification 1200
\baselineskip=12pt plus 1pt
\parindent=25pt

\font\small=cmr10 at 10truept
\baselineskip=20pt plus 1pt


{\nopagenumbers
{
\small
\baselineskip=12pt plus 1pt
\hfill Alberta-THY-25-1993

\hfill MAY 1993
}

\centerline{\bf THREE DIMENSIONAL BLACK HOLES AND FOUR DIMENSIONAL}
\centerline{\bf BLACK STRINGS AS NONLINEAR SIGMA MODELS}

\vskip1.5truecm

\centerline{{\bf Nemanja Kaloper}$~^{*}$}
{
\small
\baselineskip=12pt plus 1pt
\footnote{}{$^{*}~~${\it based on a poster presented in absentio at
the 5th Canadian Conference on General Relativity and Relativistic
Astrophysics,
Waterloo, Ontario, May 13-15, 1993, and a talk presented
at the Conference on Quantum Aspects of Black Holes, U. of California,
Santa Barbara, Califonia, June 21-27, 1993.}
 }
}
\centerline{Theoretical Physics Institute}
\centerline{Department of Physics, University of Alberta}
\centerline{Edmonton, Alberta T6G 2J1, Canada }
\vskip1.5cm
\centerline{\bf Abstract}

{
Two solutions of stringy gravity in three and
four dimensions which
admit interpretation as a black hole and a black
string, respectively, are
discussed. It is demonstrated that they are exact
WZWN nonlinear sigma models to all orders in the
inverse string tension, and hence represent
exact conformal field theories on the world-sheet.
Furthermore, since the dilaton
for these two solutions is constant, they also
solve the equations of motion of
standard GR with a minimally coupled three form field strength.
}
\vskip2cm
\vfil
\eject}

Theory of gravity has remained one of
the most challenging problems
of physics of our time. The present status
of gravity is in many ways
equivocal. Whereas in the classical domain
it is described exceptionally well
by Einstein's theory of General Relativity (GR)
all attempts to construct a consistent
quantum theory have been foiled with grave
difficulties of both technical
and conceptual nature. String theory is one of
those attempts, which
technically looks extremely attractive,
particularly for the reason of its well
behaved ultraviolet regime. It still remains to
be seen, however, what
are the basic principles of string theory,
playing the role of its cornerstone,
much the same way as the Principle
of Equivalence stands in GR.

So far, string theory has lead to
particularly  fruitful developments in the study of the
gravitational sector. Perhaps one of
the most important recent achievements was
the suggestion how string theory might be
able to avoid singularity problems
which plague many GR solutions, such as black holes.
The extra symmetries present in string theory
provide stringent constraints on the behavior of
exact solutions, and lead to a host of nonrenormalization
theorems. These could be employed to construct
exact nonsingular solutions to all orders in
both the inverse string tension and genus expansions.
In this review I will reflect on two examples,
which can be viewed as a black hole in three
and a black string in
four dimensions. They
do not change even
after all higher order corrections from the
inverse string tension expansion
(equivalently, an expansion in powers of
curvature) are included. Thence the two solutions
represent exact conformal field theories on
the world-sheet. Both
have finite curvature everywhere, except
at the origin,
and hence represent structures with horizon
but with controllable divergence.

The dynamics of the bosonic zero mass sector of
effective string theory in $D$ dimensions
is described by the effective action,
in the world sheet frame and to order $O(\alpha'^{0})$,
$$
S~=~\int d^{D}x\sqrt{G}  e^{-\sqrt{2}\kappa \Phi}
\big({1 \over 2\kappa^{2}}R -
H_{\mu\nu\lambda}H^{\mu\nu\lambda} +
\partial_{\mu}\Phi \partial^{\mu} \Phi + \Lambda \big)
\eqno(1)
$$
\eject
\noindent Here
$~H_{\mu\nu\lambda}=\partial_{[\lambda}B_{\mu\nu]}~$
is the field strength associated with the Kalb-Ramond
field $~B_{\mu\nu}~$
and $~\Phi~$
is the dilaton field.
Braces denote antisymmetrization over enclosed indices.
The cosmological constant has been included to represent the
central charge deficit
$~\Lambda={2 \over 3}\delta c_{T} = {2 \over 3}(c_{T}-D) \ge 0~$.
It arises
as the difference of the
internal theory central charge and the total central charge for a
conformally invariant theory $~ c_{tot}=26~$. Note that in the conventions
adopted here positive $\Lambda$ corresponds to a negative cosmological
constant.

The three-dimensional black hole solution is
the extension of the recent
construction of Banados, Teitelboim and
Zanelli$^{1)}$ (BTZ) into the framework
of string theory$^{2)}$. It is incorporated in
string theory by the addition of the
Kalb-Ramond axion, which in this case is
completely determined by the
third cohomology group probed by the three-form
axion field strength $H_{\mu\nu\lambda}$. The
dilaton is, surprisingly, constant due to the contribution
of the axion to the dilaton field
equation which cancels the cosmological constant.
This solution can be formulated as a nonlinear sigma
model on the world-sheet. To show it, one needs to recall the
Polyakov action for a nonlinear sigma model on the world-sheet.
It is given by ($2\sqrt{2/3}$ arises from normalizations
of the wedge product)
$$
S_{\sigma}~={1 \over \pi}
{}~\int d^2\sigma (G_{\mu\nu} + 2\sqrt{2 \over 3}
B_{\mu\nu})~\partial_{+}X^{\mu}
\partial_{-}X^{\nu}
\eqno(2)
$$
\noindent where the metric $G_{\mu\nu}$ and the axion field
$B_{\mu\nu}$ play the role of the effective coupling constants of the $2D$
field theory defined by (2). In general, there exists plethora of
various constructions which lead to a dynamical theory described by (2).
One such approach is the Wess-Zumino-Witten-Novikov (WZWN) conformal field
theory, which has first arisen in the study of non-abelian bosonization
in two dimensions. The WZWN
nonlinear sigma model action is defined by
$$
S_{\sigma}~=~{k \over 4 \pi} \int d^2\sigma
Tr\bigl(g^{-1}\partial_{+}g\,g^{-1}\partial_{-}g\Bigr)
- {k \over 12 \pi}~\int_{M} d^3 \zeta
Tr\Bigl( g^{-1}dg \wedge g^{-1}dg \wedge g^{-1}dg \Bigr)
\eqno(3)
$$

\noindent where $g$ is an element of some group $G$, and
$k$ is the (positive integer) level of the associated
Ka\v c-Moody algebra. The action above has
a very big global invariance, the continuous part of
which is $G\times G$.

One way to construct the string solutions of this
theory, which can be put in form (2), is
choosing a group $G$, the parameter space of which
represents the target manifold, and
maintaining conformal invariance. The other may be
to identify a part of the
parameter manifold by locally factoring out a subgroup of the
global invariance group $G \times G$$^{3)}$. This is accomplished with
choosing an anomaly-free subgroup $H \subset G \times G$
and gauging it
with stationary gauge fields.
Either way, after the group has been parametrized,
(3) can be rewritten in terms of the parameters in the
form (2) and the metric
and the axion are just simply read off from the resulting
expressions. The
dilaton then can be computed from the
effective action (1), as has been
mentioned above. Its appearance owes to the
requirement of conformal
invariance.

The stringy version of the BTZ black hole can be constructed either as
a sigma model on the group $SL(2,R) /P$ or on an
extremally axially gauged coset
$(SL(2,R) \times R)/(R \times P)$. The group $P$ is a discrete subgroup of
$SL(2,R)$ generated by the angular Killing vector of the metric, and is
isomorphic with $Z$. It appears as means of identification of the angle
$\phi + 2n\pi \rightarrow \phi$.The group $SL(2,R) \times R$ can be
parametrized as (with $~ab + uv = 1~$)
$$
 g~=~\left(\matrix{~a~&~u~\cr
 -v~&~b~\cr}\right)~\exp{({q \over \sqrt{k}}\theta')}
\eqno(4)
$$
The central charge of this target for the level $~k~$ is
$c_{T}=3k / (k-2) + 1 - 1$, where $\pm 1$ correspond to the
free boson and the gauging, respectively. Hence, $c_{T}=3k / (k-2)$ and
the cosmological constant is $\Lambda = 4/k$.
The gauge transformations from
the axial subgroup of $~SL(2,R) \times SL(2,R)~$ mixed with
translations along the free boson are
$$
\delta a = 2\epsilon a ~~~~\delta b = - 2\epsilon b
{}~~~~\delta u = \delta v = 0
{}~~~~\delta \theta' = {2 \sqrt{2} \over q}\epsilon c
{}~~~~\delta A_{j} = -\partial_{j}
\epsilon
\eqno(5)
$$
The remaining steps of the procedure for obtaining the solution
are to fix the gauge of the group choosing $b = \pm a$ so that the anomaly
cancels,
integrate out the gauge
fields, rescale $~\theta' \rightarrow (2c/ \sqrt{k}) ~\theta' ~$ and
take the limit $~c \rightarrow \infty~$ which effectively decouples
the  $~SL(2,R)~$ part from the gauge fields.
The gauged form of the sigma model (3) is
$$\eqalign{
S_{\sigma}(g, A)~=~S_{\sigma}(g)
&+ {k \over 2 \pi} \int d^2\sigma A_{+}
\Bigl( b\partial_{-}a - a\partial_{-}b
- u\partial_{-}v + v\partial_{-}u
+ {4 qc \over \sqrt{2}k} \partial_{-}\theta' \Bigr) \cr
&+ {k \over 2 \pi} \int d^2\sigma A_{-}
\Bigl( b\partial_{+}a - a\partial_{+}b
- v\partial_{+}u + u\partial_{+}v
+ {4 qc \over \sqrt{2}k} \partial_{+}\theta' \Bigr) \cr
&+ {k \over 2 \pi} \int d^2\sigma 4 A_{+}A_{-}
\Bigl( 1 + {2 c^2 \over k} - uv \Bigr)
\cr}
\eqno(6)
$$
where (the Wess-Zumino term vanishes by gaguge fixing)
$$
S_{\sigma}~=~-{k \over 4 \pi} \int d^2\sigma
\Bigl(\partial_{+}u \partial_{-}v
+ \partial_{-}u \partial_{+}v + \partial_{+}a \partial_{-}b
 + \partial_{-}a \partial_{+}b \Bigr)
+ {q^2 \over 2\pi} \int d^2\sigma \partial_{+}\theta' \partial_{-}\theta'
\eqno(7)
$$
The resulting Polyakov sigma model action can
be rewritten as
$$\eqalign{
S_{\sigma ~eff}~=~&-{k \over 8 \pi} \int d^2\sigma
{v^2 \partial_{+}u \partial_{-}u + u^2 \partial_{-}v \partial_{+}v +
(2 - uv)\bigl(\partial_{+}u \partial_{-}v + \partial_{-}u \partial_{+}v \bigr)
\over (1 - uv)} \cr
&+ {q^2 \over 2\pi} \int d^2\sigma \Bigl(
2 (1 - uv) \partial_{+}\theta' \partial_{-}\theta' \Bigr) \cr
&+ {q\sqrt{k} \over 2\sqrt{2}\pi} \int d^2\sigma
\Bigl( \bigl( u\partial_{-}v - v\partial_{-}u \bigr) \partial_{+}\theta' +
\bigl( v\partial_{+}u - u\partial_{+}v \bigr) \partial_{-}\theta' \Bigr) \cr}
\eqno(8)
$$
To extract the solution from (8), one needs to introduce a set of
coordinate transformations, which recast (8) into the form of the $3D$
black hole. The first transformation is
$~u~=~\exp{(\sqrt{2\over k}qt')} \sqrt{(R/q)^2 - 1 }$,
$~v~=~-\exp{(-\sqrt{2\over k}qt')} \sqrt{(R/q)^2 - 1 }$.
To introduce the
angular momentum, one can further "boost" the $t', \theta'$
coordinates to the new frame $t,\theta$ and identify along the orbits
of $\zeta = \partial /\partial \theta$. This determines the structure
of the group $P$ introduced above: $P = \exp(2n\pi\zeta)$, with n integers.
The boost is performed according to
$x^k = \tilde O^k{}_{j}x'^j$ where $\tilde O$ is an $SO(1,1)$ Lorentz
transformation, and its parameter is defined by
$$
\sinh \beta = {\rm sign}(J) {1 \over \sqrt{2}} \Bigl(
{1 - \sqrt{ 1 - (J/M)^2} \over \sqrt{ 1 - (J/M)^2} }\Bigr)^{1/2}
\eqno(9)
$$
\noindent With more definitions of the parameters,
$\rho_+^2 = q^2 \sqrt{2\Lambda} = M(1 - (J/M)^2)^{1/2}$,
$R^2 = (\sqrt{2\Lambda}/2)\bigl(\rho^2 + M - \rho_+^2 \bigr)$,
$N^{\theta} = - J /2R^2$ and $\Lambda = 4/k$,
the final solution is
$$\eqalign{
&ds^{2} =  { d\rho^{2} \over 2\Lambda (\rho^2 - \rho^2_{+})} +
{}~R^2 (d\theta + N^{\theta} dt)^2
- {\rho^2 \over R^2} {\rho^2 - \rho^2_{+} \over 2\Lambda} dt^2 \cr
&~~~~~~~~~~~~~~~~~~~~~~~~~~~~~~
{}~~B_{t\theta} = {\rho^2 \over \sqrt{6\Lambda}}
}
\eqno(10)
$$
\noindent The dilaton can be found from the
associated  effective action,
or from a careful computation of the
Jacobian determinant arising from integrating out the gauge fields.
Inspection of the Jacobian
before the limit $~c \rightarrow \infty~$ is taken gives
$~J \propto 1/(1 + (2c^2 /k) - uv)~$
$= (k/2c^2)/\bigl(1 + (k/2c^2)(1 - uv)\bigr)~$.
As $~c \rightarrow \infty~$ the non-constant terms decouple and do not
contribute to the dilaton. Thus $\Phi = \Phi_0 = {\rm const}$.

The metric part is almost precisely the BTZ solution. The only difference is,
the cosmological constant in  (10) is half that of what on obtains
in ordinary GR in three dimensions.
The reason for this discrepancy is, that the presence of
the axion introduces an extra contribution
to the cosmological constant, which
just cancels one half of it, since the dilaton is constant.
This property of the solution (10) is interesting, since the
absence of the dilaton dynamics guarantees that the solution is
also a solution of standard GR with a minimally coupled two form, as
can be immediately verified from action (1), after $\Phi = {\rm const.}$
is substituted.

The solution (10) can be immediately extended to four dimensions$^{4)}$, by
tensoring it by a flat direction. The only change in the solution (10) is
that in four dimensions there is an extra additive $dz^2$ term in the metric.
This solution can be understood as a black string in four dimensions. Its
conformal field theory representation is an extremally axially gauged coset
$(SL(2,R) \times R^2)/(R \times P)$ on the
level $k$. Interpretation of this
extension of (10) as a rotating
black string is best seen if one replaces the three
form $H_{\mu\nu\lambda}$ by its dual. The dual
axion field strength $V=\sqrt{6\Lambda}dz=da(z)$
can be integrated between any two
space-like ($t={\rm const}$) hypersurfaces $z_{1,2}={\rm const}$ to give
$a(z_2) - a(z_1) = \sqrt{6\Lambda} \Delta z$.
Therefore, the axion solution can be
understood as a constant gradient of the pseudoscalar axion field. As
$z_{1,2} \rightarrow \infty$, the axion diverges. This is easy to explain:
it is a consequence of the assumption that the string is infinitely
long. In reality, one should expect some cut-off sufficiently far away along
the string. The situation is analogous to that of the
electrostatic potential between the plates of a parallel plate capacitor in
ordinary electromagnetism. The cut-off occurs on the plates of the
capacitor, where the potential is constant. The gradient is just
$\vec \nabla V = (\Delta V/\Delta L) \vec z$. This analogy shows that the
black string solution should be viewed as a gravitational configuration
which arose inside a transitory region
separating two domains within which
the axion is constant, $a_1$ and $a_2$ respectively. The axion gradient
inside this region corresponds to the adiabatic change in the axion vacuum,
where the adiabatic approximation is better if
the transitory region (and hence the string)
is bigger. The string evidently needs
the domain of axionic gradient for
its existence (because the axion gradient stops the dilaton from rolling), and
thence can be labelled primordial. It should be noted that
in four dimensions, the axion
also plays role of a Higgs field. The axion condensate
$6\Lambda$ in (1) breaks the
normal general covariance group $GL(3,1)$ of (1) down to $GL(2,1)$.

Higher order corrections could now be investigated following the
recently established resummation procedure$^{5)}$.
It turns out, that both  configurations actually survive the
corrections, and appear to be exact solutions of string theory to all
orders in $\alpha'$. The only effect of the higher order $\alpha'$
corrections is finite renormalization of the parameters in (10), and
in particular, renormalization of the semiclassical expression for the
cosmological constant.

In summary, in this review it was shown how the complex
structure of string theory can be employed for the construction of exact
solutions which are consistent to all orders in the inverse string tension
expansion. A clear advantage of this program is that the solutions
constructed as nonlinear sigma models can be analyzed in relationship to the
exact effective action involving higher powers of curvature in a rather
elegant way. Furthermore, the two examples exibited here are found to be
exact solutions of the exact effective action, and hence may be good candidates
for consistent quantum gravitational configurations.

\vfill
\eject
\noindent {\bf Acknowledgements}
\vskip.7truecm
Thanks are due to B. Campbell, G. Hayward, V. Husain, W. Israel
and D. Page for useful conversations.
\vskip.7truecm
\noindent {\bf References}
\vskip.7truecm

\item{1)} M. Banados, C. Teitelboim, and
J. Zanelli, Phys. Rev. Lett.
{\bf 69} (1992) 1849; M. Banados,  M. Henneaux,
C. Teitelboim and J. Zanelli, IAS preprint HEP-92/81;
D. Cangemi, M. Leblanc and R.B. Mann,
MIT preprint CTP-2162, 1992; S.F. Ross and R.B. Mann,
Phys. Rev. {\bf D47} (1993) 3319.

\item{2)} G.T. Horowitz and D.L. Welch,
Phys. Rev. Lett. {\bf 71} (1993) 328;
N. Kaloper, Univ. of Alberta
preprint Alberta-Thy-8-93, Feb. 1993, in press in Phys. Rev {\bf D}.

\item{3)} E. Witten, Phys. Rev. {\bf D44} (1991) 314;
J.H. Horne and G.T. Horowitz, Nucl. Phys. {\bf B368} (1992) 444.

\item{4)} N. Kaloper, Univ. of Alberta
preprint Alberta-Thy-14-93, March 1993, in press in Phys. Rev. {\bf D}.

\item{5)} A. Tseytlin, Nucl. Phys. {\bf B399} (1993) 601;
CERN preprint CERN-TH-6804/93, Feb. 1993;
I. Bars and K. Sfetsos,
Phys. Rev. {\bf D48} (1993) 844.

\bye